# Forecasting Implied Volatility Smile Surface via Deep Learning and Attention Mechanism

Shengli Chen[1,2],   Zili Zhang [2]
1. Guanghua School of Management, Peking University, Beijing 100871
2. Harvest Fund Management Co., Ltd., Beijing 100005

**Abstract**: The implied volatility smile surface is the basis of option pricing, and the dynamic evolution of the option volatility smile surface is difficult to predict. In this paper, attention mechanism is introduced into LSTM, and a volatility surface prediction method combining deep learning and attention mechanism is pioneeringly established. LSTM's forgetting gate makes it have strong generalization ability, and its feedback structure enables it to characterize the long memory of financial volatility. The application of attention mechanism in LSTM networks can significantly enhance the ability of LSTM networks to select input features. This paper considers the discrete points of the implied volatility smile surface as an overall prediction target, extracts the daily, weekly, and monthly option implied volatility as input features and establishes a set of LSTM-Attention deep learning systems. Using the dropout mechanism in training reduces the risk of overfitting. For the prediction results, we use arbitrage-free smoothing to form the final implied volatility smile surface. This article uses the S&P 500 option market to conduct an empirical study. The research shows that the error curve of the LSTM-attention prediction system converges, and the prediction of the implied volatility surface is more accurate than other predicting system. According to the implied volatility surface of the 3-year rolling forecast, the BS formula is used to pricing the option contract, and then a time spread strategy and a butterfly spread strategy are constructed respectively. The experimental results show that the two strategies constructed using the predicted implied volatility surfaces have higher returns and Sharpe ratios than that the volatility surfaces are not predicted. This paper confirms that the use of AI to predict the implied volatility surface has theoretical and economic value. The research method provides a new reference for option pricing and strategy.

## 1 Introduction

Volatility is one of the key attributes of asset prices and a key variable in the pricing of options contracts. In the classic Black-Scholes-Merton (BSM) option pricing model, it is assumed that the volatility is constant, and the implied volatility is the asset price, exercise price, remaining period, risk-free interest rate, bonus interest rate, etc. Given those parameters, the market price of the European option is substituted into the BSM pricing formula to reverse the volatility. According to the assumption of the BSM option pricing formula, the implied volatility is a

---

**Biography:**

Shengli Chen, Ph.D. in Management, He is Postdoctoral Research Fellow at the Peking University - Harvest Fund joint postdoctoral research station. His research interests Artificial Intelligence and Quantitative Investment. His representative paper titled "Forecasting Realized volatility of Chinese Stock Index Futures based on Jumps, Good-bad Volatility and Baidu Index" was published in the ***Systems Engineering — Theory & Practice***（Issue 2, 2018) (in Chinese). E-mail: chensl@jsfund.cn.

Zili Zhang, Ph.D. in Theoretical Physics, University of Texas at Austin, Managing Director of Harvest Fund Management Co., Ltd., Head of AI Investment & Research Center, Fund Manager. His research focuses on Artificial Intelligence and Quantitative Investments. His representative paper titled "Stock Network, Systematic Risk and Stock Pricing Implications" will be published in the Quarterly Journal of Economics (forthcoming in 2020.1) (in Chinese). E-mail: zhangzl@jsfund.cn

**Fund**: The paper is funded by the First-class Project of China Postdoctoral Science Foundation (2018M640367)

constant that has nothing to do with the exercise price and the remaining period of the option. However, what people observe in the market is that the implied volatility varies with the exercise price and the remaining term. By plotting the implied volatility corresponding to options with different durations and different exercise prices in a three-dimensional space composed of terms, exercise prices, and implied volatility, a so-called "implicit volatility surface" is formed. Moreover, a large number of studies have also found that the shape of the implied volatility surface is not static, but shows obvious time-varying characteristics over time (Dumas et al., 1998; Cont et al., 2002; Fengler et al., 2007). Today, scholars generally accept the view that the implied volatility surface at a given point in time appears as a surface shape related to the strike price and the expiration date, and is dynamically changing in the time dimension (Goncalves et al., 2006).

Volatility is one of the key attributes of asset prices and a key variable in the pricing of options contracts. In the classic Black-Scholes-Merton (BSM) option pricing model, it is assumed that the volatility is constant, and the implied volatility is the asset price, strike price, and balance given the implicit volatility surface. The question to be studied in this article is: Can the dynamic change of the implied volatility surface be predicted, and how to predict it? The predictability of asset prices is an eternal theme of financial research, and the predictability of option prices is no exception. However, due to the complexity of options, it is difficult to directly study the predictability of option prices. When the parameters of the BSM option pricing model are known, the corresponding implied volatility can be calculated by giving the fixed-term price, and the implied volatility can be reversed to the option price. Therefore, the option price and the implied volatility are formed. The one-to-one correspondence naturally becomes a surrogate variable for studying the nature of option prices. If you can accurately predict the changes in implied volatility, you can effectively predict the changes in option prices, which is of great significance to options trading and risk management. In addition, the implied volatility itself contains market participants' expectations of the future volatility of the underlying asset. Research on the predictability of the implied volatility can further identify the market risk of the underlying asset. Therefore, the dynamic modeling and extrapolation of the implied volatility surface have a fundamental effect on both the option itself and the underlying asset.

In the field of implicit volatility surface prediction, there are not many research results. There are two main reasons. First, high-dimensional research is too difficult. It is obviously more difficult to study the dynamic process of the volatility surface than the dynamic process of the stock price (single point) and the dynamic process of the term structure of interest rates (a line); Confused with the prediction of "implied volatility." Historical volatility is the volatility calculated based on the underlying asset price series, such as standard deviation, idiosyncratic volatility, and realized volatility. In recent years, many scholars have established a series of methods such as GARCH models and HAR models to predict the realized volatility. The implied volatility of the options to be predicted in this article is a direct reflection of the future volatility expectations formed by the market in options trading, which is completely different from the connotation of historical volatility, which will necessarily make its properties, characteristics and research methods different. . Compared with the prediction research of historical volatility, there are few researches on the prediction of implied volatility.

In the research literature of implicit volatility surface prediction, Carr et al. (2016) used the arbitrage-free condition to derive the expression of the implicit volatility surface, describing the dynamic behavior of the implicit volatility surface, but these studies did not involve the prediction of implied volatility. Guidolin et al. (2003) and Garcia et al. (2003) theoretically analyzed the possibility of predicting implied volatility using mechanism transformation and information learning models. Harvey et al. (1992), Guo (2000), Brooks et al. (2002) respectively studied the predictability of the implied volatility of S&P100 stock index options, foreign exchange options, and government bond options, but these studies only focused on the implied volatility of short-term options rather than the entire volatility surface. In addition, Fernandes et al. (2014), Konstantinidi et al. (2008) used the volatility index VIX to study the predictability of implied volatility, which is also not for the entire volatility surface. Researches similar to this article are Dumas et al. (1998), Goncalves et al. (2006), Bernales et al. (2014). Dumas et al. (2014) first used a two-step method to construct an implicit volatility surface. In the first step, the implied volatility surface is expressed as a function of the moneyness and the remaining period; in the second step, a time series model is constructed for the coefficients of cross-sectional regression. Goncalves et al. (2006) also used a two-step method to construct a dynamic model of the implied volatility surface of the S&P500 index option, and then tested the predictability of the implied volatility from a statistical significance. Based on the two-step method to establish the implied volatility surface model, Bernales et al. (2014) studied the predictability of the implied volatility surface of US stock options, and obtained similar conclusions as the above study.

Advances in artificial intelligence technology have provided new methods for financial engineering research. Hochreiter (1997) first proposed long short-term memory (LSTM). With the rise of deep learning, LSTM has a large number of applications in engineering fields such as natural language processing and language recognition. The LSTM has been cited in the prediction of financial time series for several reasons: 1) The feedback structure of the LSTM enables it to characterize the long memory effect of financial time series, especially for the volatility of financial assets; 2) LSTM has a unique forgetting gate, which reduces the risk of over-fitting the neural network and makes it have stronger out-of-sample prediction capabilities; 3) LSTM is an open prediction framework that can adapt Adjusting the weight of the input features on the ground has a stronger mode approximation ability than traditional linear models. In addition, in the field of machine translation, the Google team proposed an attention model that simulates the human brain (Ashish et al, 2017). The attention mechanism is mainly used to simulate the human perspective system, and jumps to different objects and positions in the entire field of view, so as to perform analysis and processing resource allocation and coordination. When human eyes look at a painting, they often focus on a small piece. At this time, the corresponding weight of the Attention model becomes larger. Introducing the attention mechanism into the LSTM structure can significantly enhance the automatic screening and optimization capabilities of LSTM input features. At present, there is no financial literature that uses LSTM and attention models for the modeling and forecasting of implicit volatility surfaces. This article will turn on artificial intelligence research methods are of great significance for modeling and forecasting of implicit volatility surfaces of options.

The structure of this paper is as follows: Section 1 is the introduction of the paper; Section 2 describes the theoretical research methods of this paper, including option pricing and implicit volatility theory, and the attention-based LSTM deep learning prediction system; Section 3 expands on Empirical research, including data sources, data, prediction results of deep learning models, and analysis of options spread strategies. Section 4 summarizes and looks forward.

## 2. Theoretical methods

### 2.1 Option pricing and implied volatility

This section first describes the landmark BS option pricing formula. In the field of option pricing research, scholars Black, Scholes, and Merton pioneered the BS model (or the BSM model) and were awarded the Nobel Prize in Economics. Under many necessary assumptions, Black, Scholes, and Merton established the following famous B-S-M differential equations:

$$\frac{\partial f}{\partial t} + rS\frac{\partial f}{\partial s} + \frac{1}{2}\frac{\partial^2 f}{\partial s^2}\sigma^2 s^2 = rf \qquad (1)$$

Through the derivation of the above differential equation, we finally get the BS pricing formula for call and put options:

$$C = S_0 \text{N}(d_1) - Ke^{-rT} \text{N}(d_2) \tag{2}$$

$$P = Ke^{-rT} \text{N}(-d_2) - S_0 \text{N}(-d_1) \tag{3}$$

$$d_1 = \frac{\ln(S_0/K) + (r + \sigma^2/2)T}{\sigma\sqrt{T}} \tag{4}$$

$$d_2 = \frac{\ln(S_0/K) + (r - \sigma^2/2)T}{\sigma\sqrt{T}} \tag{5}$$

In the above formula, $C$ represents the price of call options; $P$ represents the price of put options; $S_0$ is the starting price of the underlying asset; $r$ is the risk-free interest rate of continuous compounding; $\sigma$ is the volatility of the underlying asset; $T$ is the term of the option; $\text{N}(\cdot)$ is the cumulative probability distribution function of the standard normal distribution; $d_1$ describes the sensitivity of the option to the underlying price, and $d_2$ describes the possibility of the option being finally executed. Therefore, the theoretical prices of call and put options can be expressed by the BS formula as expressions of starting price, exercise price, risk-free interest rate, asset volatility and option duration.

Implied volatility is a core variable in option pricing and investment trading. Implied volatility refers to the implied volatility of the market price of an option. The most common method for estimating implied volatility is to substitute the market price of options into the BS formula and calculate implied volatility backwards. In other words, according to the BS formula, we can calculate the implied volatility at a certain moment at a specific strike price. As the option exercise price changes, the implied volatility calculated by the BS formula presents a smile curve, that is, the volatility smile curve, which describes the relationship between the implied volatility of the option and the exercise price function. On this basis, if the change of the term is further considered, the implied volatility calculated by the BS formula will form a three-dimensional volatility smile surface. In actual options trading, investors often use the volatility smile curve and volatility smile surface to analyze and find trading opportunities.

Because the exercise price and term of the option contract are discrete, the implied volatility smile surface calculated by using the BS formula to calculate backwards point by point is also discretized and presents an uneven surface. On this basis, academia and practice circles usually perform non-arbitrage smoothing, and finally get a smoother volatility smile surface. The real significance of the predicted volatility smile surface is that the next period of option contracts can be priced based on the predicted volatility surface, thereby laying the foundation for the study of option strategies.

**2.2 Recurrent neural networks**

Recurrent neural networks (RNNs, also known as "feedback neural networks") are an extension of feed-forward neural networks, which include feedback hidden layers (recurrent hidden states), whose output in the current state depends on the value of the previous state. Given the input vector of the neural network:

$$\mathbf{x} = (\mathbf{x}_1, \mathbf{x}_2, \cdots, \mathbf{x}_t, \cdots, \mathbf{x}_n) \tag{6}$$

The recurrent neural network updates its feedback hidden layer vector by the formula:

$$\mathbf{h}_t = \phi(\mathbf{W}\mathbf{x}_t + \mathbf{U}\mathbf{h}_{t-1} + \mathbf{b}) \tag{7}$$

where the note **b** represents the deviation vector and $\phi$ represents the activation function of the feedback hidden layer. The expanded structure of a recurrent neural network is shown in Figure 1. The figure shows that the recurrent neural network in the feedback hidden layer $A(t)$ at state $t$ contains the input vector $\mathbf{x}_t$ and the feedback hidden layer output vector $\mathbf{h}_{t-1}$ of the previous state, and calculates a new output vector $\mathbf{h}_t$. In this process, the state of $A(t)$ is passed from the previous step to the next step. Therefore, a recurrent neural network is theoretically the result of a finite number of replications of the same neural network structure. At present, recurrent neural networks have been widely used in the research of cutting-edge issues such as speech recognition, language models, machine translation, and timing analysis, with great success. In the field of financial market predictive modeling, Danilo and Jonathon et al. (2002) systematically studied the theoretical advantages, modeling methods, and economic value of RNNs feedback loop structure.

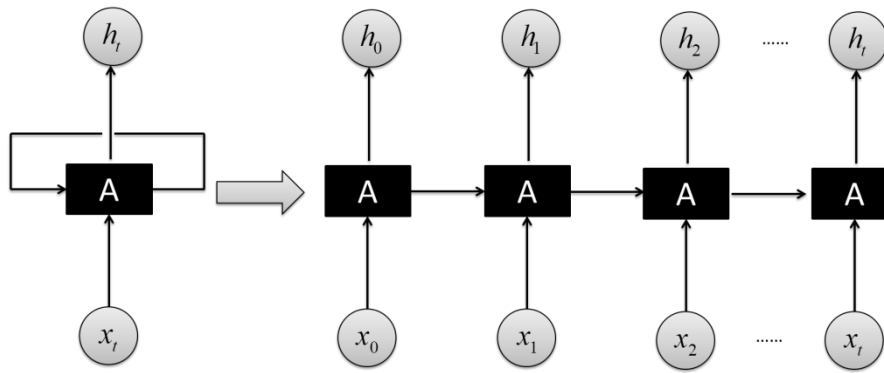

Fig.1 The unfolding structure of recurrent neural network

**2.3 Long Short-Term Memory**

The working mechanism of recurrent neural networks is to use historical information to assist current decisions. It can better use information that cannot be modeled by feedforward neural network structures, but it also brings long-term dependencies to the technology challenge. In many practical prediction problems, useful information that affects the current state can vary in time from the current state to different lengths. If the time interval of useful information is short, the training results of RNNs are susceptible to interference by historical redundant information; if the interval of useful information is long, the loop iteration of RNNs increases a large number of calculations and it is difficult to capture useful key information. The design of long short-term memory (LSTM) can effectively solve the above problems faced by RNNs. As a specially designed feedback neural network structure, LSTM performs better than standard RNNs on many tasks and is one of the most successful representatives of RNNs engineering practice.

The earliest LSTM was proposed by Hochreiter et al. (1997), which replaced the RNN hidden layer with a memory block and used a memory cell to store information. In order to enhance the processing capability of LSTM, Gers et al. (2000) introduced forget gate. Gers et al. (2000, 2002) further proposed peephole connections on this basis, which added the cell states of the previous state as inputs for three gate structures. Therefore, the current popular LSTM network structure is shown in Figure 2. The core of the LSTM network structure is three special gate structures: input gate, output gate, and forget gate.

They are composed of weighted multipliers and sigmoid activation functions. They can predict the output at a certain moment by remembering or forgetting the accumulated information in the range [0,1]. The forget gate determines which part of the information needs to be forgotten according to the input $\mathbf{x}_t$ of the current state, $\mathbf{c}_{t-1}$ of the previous state, and output $\mathbf{h}_{t-1}$ of the previous state. The input gate decides which part of the information enters the state $\mathbf{c}_t$ at the current moment according to $\mathbf{x}_t$、 $\mathbf{c}_{t-1}$ and $\mathbf{h}_{t-1}$. Therefore, the gate structure of LSTM can more effectively decide which information is forgotten and which information should be retained.

With reference to Gers et al. (2000, 2002) and Alex (2014), the calculation formula of LSTM network is:

$$\mathbf{i}_t = \sigma(\mathbf{W}_{ix}\mathbf{x}_t + \mathbf{W}_{ih}\mathbf{h}_{t-1} + \mathbf{W}_{ic}\mathbf{c}_{t-1} + \mathbf{b}_i) \qquad (8)$$

$$\mathbf{f}_t = \sigma(\mathbf{W}_{fx}\mathbf{x}_t + \mathbf{W}_{fh}\mathbf{h}_{t-1} + \mathbf{W}_{fc}\mathbf{c}_{t-1} + \mathbf{b}_f) \qquad (9)$$

$$\mathbf{o}_t = \sigma(\mathbf{W}_{ox}\mathbf{x}_t + \mathbf{W}_{oh}\mathbf{h}_{t-1} + \mathbf{W}_{oc}\mathbf{c}_{t-1} + \mathbf{b}_o) \qquad (10)$$

$$\mathbf{g}_t = \phi(\mathbf{W}_{gx}\mathbf{x}_t + \mathbf{W}_{gh}\mathbf{h}_{t-1} + \mathbf{b}_g) \qquad (11)$$

$$\mathbf{c}_t = \mathbf{f}_t \odot \mathbf{c}_{t-1} + \mathbf{i}_t \odot \mathbf{g}_t \qquad (12)$$

$$\mathbf{h}_t = \mathbf{o}_t \odot \phi(\mathbf{c}_t) \qquad (13)$$

Where $\sigma, \phi$ are activation functions, respectively representing sigmoid function and tanh function; $\odot$ is element-wise multiplication; $\mathbf{W}$ is weight matrix; $\mathbf{b}$ is error vector; $\mathbf{i}_t$、$\mathbf{f}_t$ and $\mathbf{o}_t$ represent input gate, forget gate, and output gate at time t; $\mathbf{c}_t$ represents an iterative method of cell states; $\mathbf{h}_t$ represents the network output. As can be seen in Figure 2, the input gate, forget gate, and output gate are all connected to the point multiplication node, which respectively represent the states of the input, output, and memory units that control the LSTM network. The training algorithms for LSTM can be read in Hochreiter et al. (1997) and Gers et al. (2002), which are not described in this article.

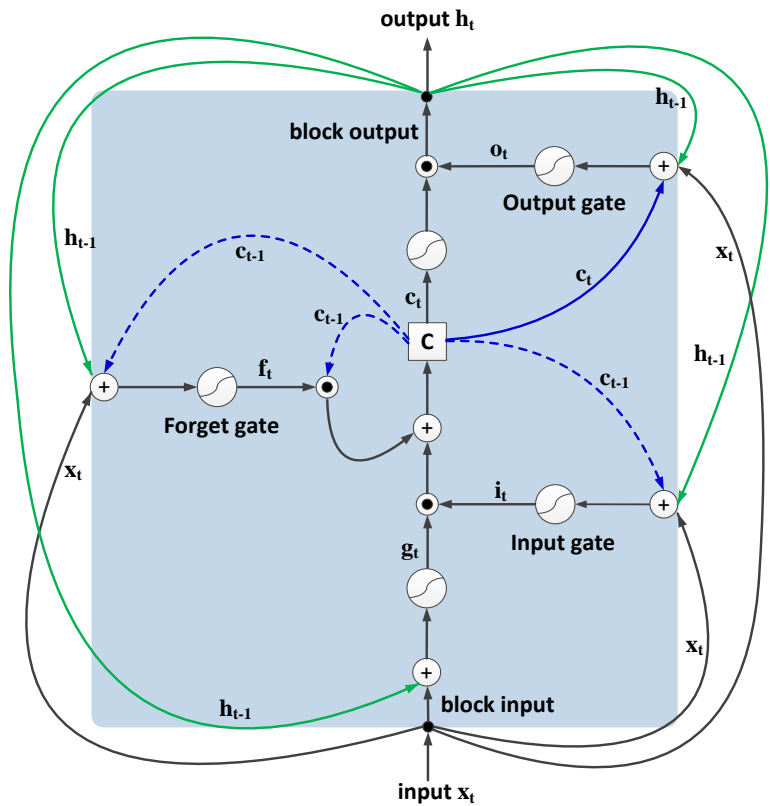

Fig.2 The structure of long short-term memory network

## 2.4 Attention mechanism

The Google team first proposed an Attention model that simulates human visual attention mechanism (Ashish et al, 2017). In recent years, attention models have been widely used in various fields of deep learning. Whether it is image processing, speech recognition or natural language processing, it is often encountered in various types of tasks. The visual attention mechanism is a brain signal processing mechanism unique to human vision. Human vision quickly scans the global image to obtain the target area that needs attention, and then invests more attention resources in this area to obtain more detailed information about the target that needs attention, and suppress other useless information. The human visual attention mechanism is a means for humans to quickly screen out high-value information from a large amount of information using limited attention resources, which greatly improves the efficiency and accuracy of visual information processing. The attention mechanism in deep learning is similar to the human visual attention mechanism in essence. The core goal is to select information that is more critical to the current task goal from a lot of information.

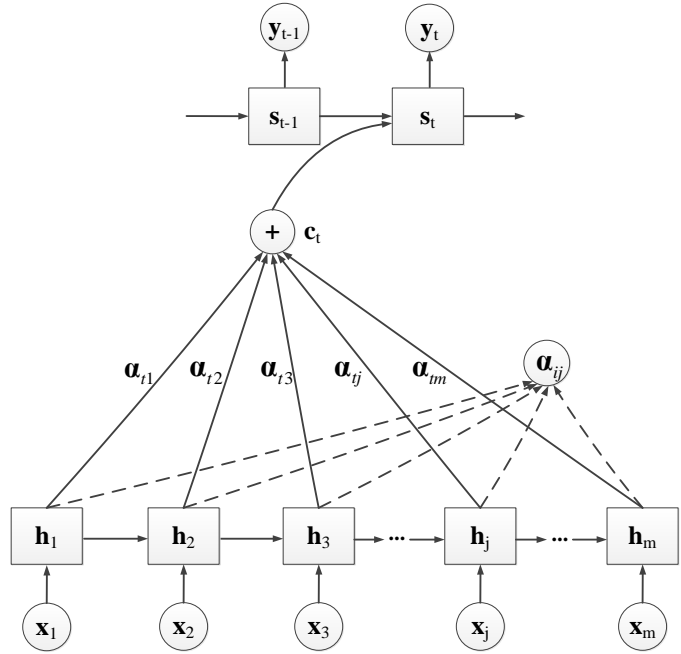

Fig.3 The graphical illustration of attention mechanism for LSTM network

The method of increasing the attention mechanism in the LSTM network is to embed an attention model between the two LSTM hidden layers, so that the hidden state affecting the prediction target is automatically assigned more attention weights, thereby optimizing and enhancing the LSTM's ability to learn key features. Figure 3 introduces the attention mechanism of two adjacent LSTM hidden layers, where $\mathbf{x}=(\mathbf{x}_1,\mathbf{x}_2,\cdots,\mathbf{x}_m)$ represents the input of the first LSTM hidden layer; $\mathbf{h}=(\mathbf{h}_1,\mathbf{h}_2,\cdots,\mathbf{h}_m)$ represents the hidden state of the LSTM; $\mathbf{s}=(\mathbf{s}_1,\mathbf{s}_2,\cdots,\mathbf{s}_n)$ is the state of the next LSTM hidden layer, $\mathbf{y}=(\mathbf{y}_1,\mathbf{y}_2,\cdots,\mathbf{y}_n)$ is the corresponding prediction targets; $\mathbf{c}_t$ is the memory unit of the attention model, and $\boldsymbol{\alpha}_t=(\boldsymbol{\alpha}_{t1},\boldsymbol{\alpha}_{t2},\cdots,\boldsymbol{\alpha}_{tm})$ is the corresponding attention weight allocation vector.

The core of understanding the attention model is to sort out the calculation of current hidden state $\mathbf{s}_t$. Let $\mathbf{s}_t$ be the hidden layer state of step LSTM, then:

$$\mathbf{s}_t = f(\mathbf{s}_{t-1},\mathbf{y}_{t-1},\mathbf{c}_t) \tag{14}$$

Where $\mathbf{y}_{t-1}$ is the output target of the previous step; $\mathbf{c}_t$ is the memory unit, and its calculation formula is:

$$\mathbf{c}_t = \sum_{j=1}^{m}\boldsymbol{\alpha}_{tj}\mathbf{h}_j \tag{15}$$

Where $\mathbf{h}_j$ is the LSTM hidden layer state corresponding to the input $\mathbf{x}_j$ of step $j$; $\boldsymbol{\alpha}_{tj}$ is the attention distribution weight of memory unit $\boldsymbol{\alpha}_{tj}$ on hidden layer state $\mathbf{h}_j$, and its calculation formula is:

$$\boldsymbol{\alpha}_{tj} = \frac{\exp(e_{tj})}{\sum_{k=1}^{m} \exp(e_{tk})} \tag{16}$$

where $e_{tj}$ is an alignment model, which is a non-linear approximation function $e_{tj}=a(\mathbf{s}_{t-1},\mathbf{h}_j)$ of the states $\mathbf{s}_{t-1}$ and $\mathbf{h}_j$ of the two hidden layers before and after the LSTM. To reduce the amount of calculation, this paper uses a single-layer perceptron as the approximation function:

$$a(\mathbf{s}_{t-1},\mathbf{h}_j) = \lambda \cdot \tanh(\mathbf{W}_{ah}\mathbf{h}_j + \mathbf{W}_{as}\mathbf{s}_{t-1} + \mathbf{b}_a) \tag{17}$$

In summary, the attention model takes the hidden layer state $\mathbf{h}=(\mathbf{h}_1,\mathbf{h}_2,\cdots,\mathbf{h}_m)$ of the LSTM as input, and after calculation and processing of the attention layer (Attention layer), it finally maps to the LSTM hidden layer state $\mathbf{s}=(\mathbf{s}_1,\mathbf{s}_2,\cdots,\mathbf{s}_n)$.

**2.5 Dropout training mechanism**

In deep learning predictive modeling, how to avoid overfitting is a key point in the application of neural networks. The current effective way is to add a dropout training mechanism to the prediction system. The earliest dropout training method was first proposed by Hinton et al. (2014), and Yarin (2015) subsequently introduced it to the training of RNNs. The Dropout training method is mainly used in neural network training to discard part of the data of the memory unit with a specific probability, thereby reducing the overfitting of the neural network within the sample and enhancing the generalization ability outside the sample. The Dropout training mechanism can be used both inside the LSTM memory unit (Yarin, 2015) and outside the LSTM hidden layer. In this paper, the use of the dropout training mechanism is mainly to modify the state output $\mathbf{h}$ of the hidden layer of the LSTM in network training.

Drawing on Hinton et al. (2014), this article explains the mathematical principles of the dropout training mechanism. Let $l \in \{1,\cdots,L\}$ be the number of the hidden layer of the neural network, $\mathbf{h}^{(l)}$ be the output vector of the l-th hidden layer, $\mathbf{w}^{(l)}$ and $\mathbf{b}^{(l)}$ be the weight vector and the bias vector. As shown in Figure 4 (a), the calculation formula for the neuron node $i$ is expressed as:

$$\begin{aligned} z_i^{(l+1)} &= \mathbf{w}_i^{(l+1)}\mathbf{h}^{(l)} + b_i^{(l+1)} \\ y_i^{(l+1)} &= f(z_i^{(l+1)}) \end{aligned} \tag{18}$$

Where $z_i^{(l+1)}$ is the weighted sum of the input features and $f$ is the activation function. As shown in Figure 4 (b), when the input data of this neuron node uses the dropout training mechanism, its calculation formula is:

$$\begin{aligned} r_j^{(l)} &\sim \text{Bernoulli}(p) \\ \tilde{\mathbf{h}}^{(l)} &= \mathbf{r}^{(l)}\mathbf{h}^{(l)} \\ z_i^{(l+1)} &= \mathbf{w}_i\tilde{\mathbf{h}}^{(l)} + b_i^{(l+1)} \\ y_i^{(l+1)} &= f(z_i^{(l+1)}) \end{aligned} \tag{19}$$

where $\mathbf{r}^{(l)}$ is a Bernoulli independent and identically distributed vector, and the probability of each element being 1 is p (the probability of 0 is 1-p); $\tilde{\mathbf{h}}^{(l)}$ is the hidden layer output vector after dropout probability correction. It is worth mentioning that the dropout mechanism is only used for the training of neural networks. In actual prediction, the neuron node i is calculated according to the non-dropout formula.

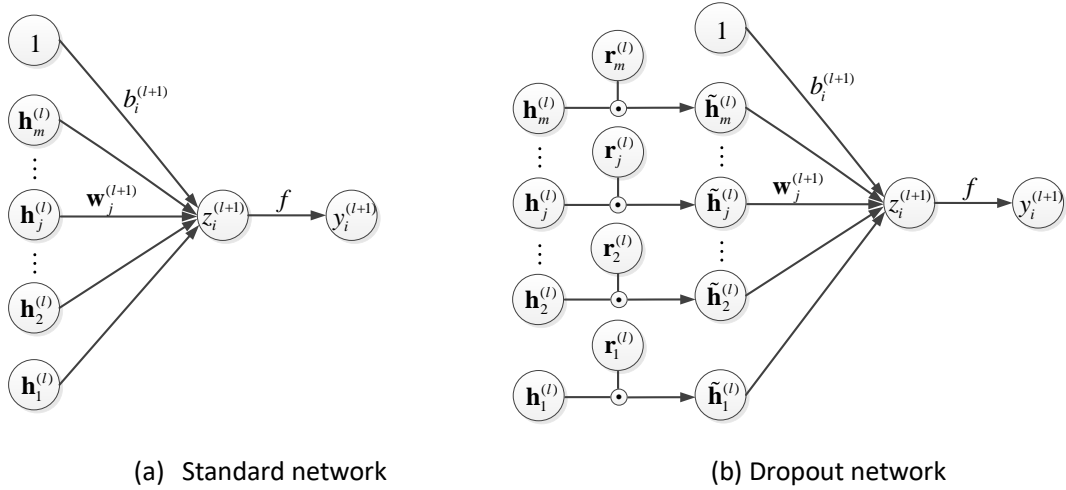

(a) Standard network  (b) Dropout network

Fig.4 The basic operations of a standard and dropout network

## 2.6 Deep Learning Prediction System

The theory of LSTM memory unit, attention model, and dropout training mechanism was explained in the previous article, but they are only local systems that predict the volatility smile surface. A complete deep learning prediction system needs to organically combine them. This paper builds an attention-based LSTM deep learning prediction system for the volatility smile surface prediction, which can be regarded as a deep neural network model (abbreviated as Att-LSTM) for volatility surface prediction. If the input feature of the prediction system is $\mathbf{x}_T$, and the prediction target is $\mathbf{y}_{T+1}$, then the mathematical form of the deep learning prediction system constructed in this paper is $\mathbf{y}_{T+1}=\text{Att-LSTM}(\mathbf{x}_T)$. The network structure of the Att-LSTM deep learning prediction system is shown in Figure 5. From bottom to top, the prediction system is input layer, LSTM hidden layer, dropout training layer, attention layer, LSTM hidden layer, dropout training layer, fully connected layer and output layer. Therefore, the prediction system includes 2 LSTM hidden layers, 2 dropout training layers, and 1 attention layer.

The information transfer process of the prediction system includes the following core links. 1) Input layer. For the prediction target $\mathbf{y}_{T+1}$, if the external input of the prediction system is a matrix of dimension T*N, then the input layer is composed of the feature vectors of adjacent T steps $\mathbf{x}_T=(x_1, x_2, \cdots, x_T)$, and each step input is a vector of N input features, that $\mathbf{x}_j = (x_{j1}, x_{j2}, \cdots, x_{jN})$. 2) The first LSTM hidden layer. Each step of the input feature $\mathbf{x}_j$ will correspond to a memory unit in the hidden layer of LSTM. Its internal structure is shown in Figure 2. The state output vector $\mathbf{h}_j^{(1)}$ of the memory unit is passed to the next layer as the LSTM output result on the one hand, and passed to the next memory unit on the other hand. 3) The first dropout layer. During the LSTM training process, the Dropout layer modifies the hidden layer state state $\mathbf{h}_j$ with a specific probability to $\tilde{\mathbf{h}}_j$ to increase the generalization ability of the prediction system. 4) Attention layer. The vector $\tilde{\mathbf{h}}_j$ is converted into a vector $\mathbf{h}_j^{(2)}$ according to the attention distribution weight, and is used as an input vector of the second-level LSTM memory unit. 5) The second LSTM hidden layer. The difference between this LSTM hidden layer and the first LSTM hidden layer is that it only uses the state of the last step LSTM memory unit as the output of the entire LSTM hidden layer. 6) The second dropout layer. First map the LSTM state vector to N neurons, and add a dropout training mechanism to the state value $\mathbf{h}_j^{(3)}$ of each neuron. 7) Fully connected layer. The fully connected layer is a simple perceptron, which weights and sums the vector $\tilde{\mathbf{h}}_j^{(3)}$ after dropout processing. 8) Output layer. Select the activation function f and map the input of the ANN to the prediction target vector $\mathbf{y}_{T+1}$. It must be clear that the dimensions $\mathbf{h}_j^{(1)}$ and $\mathbf{h}_j^{(3)}$ of the output vector of the LSTM hidden layer in Figure 5 can be flexibly defined. The dimensions of the output vector $\mathbf{y}_{T+1}$ of the fully connected layer need to match the prediction target, while the output vectors of the dropout layer and attention layer remain the same. Regarding the information transmission process of attention-based LSTM, this article will also elaborate on it in the subsequent empirical research.

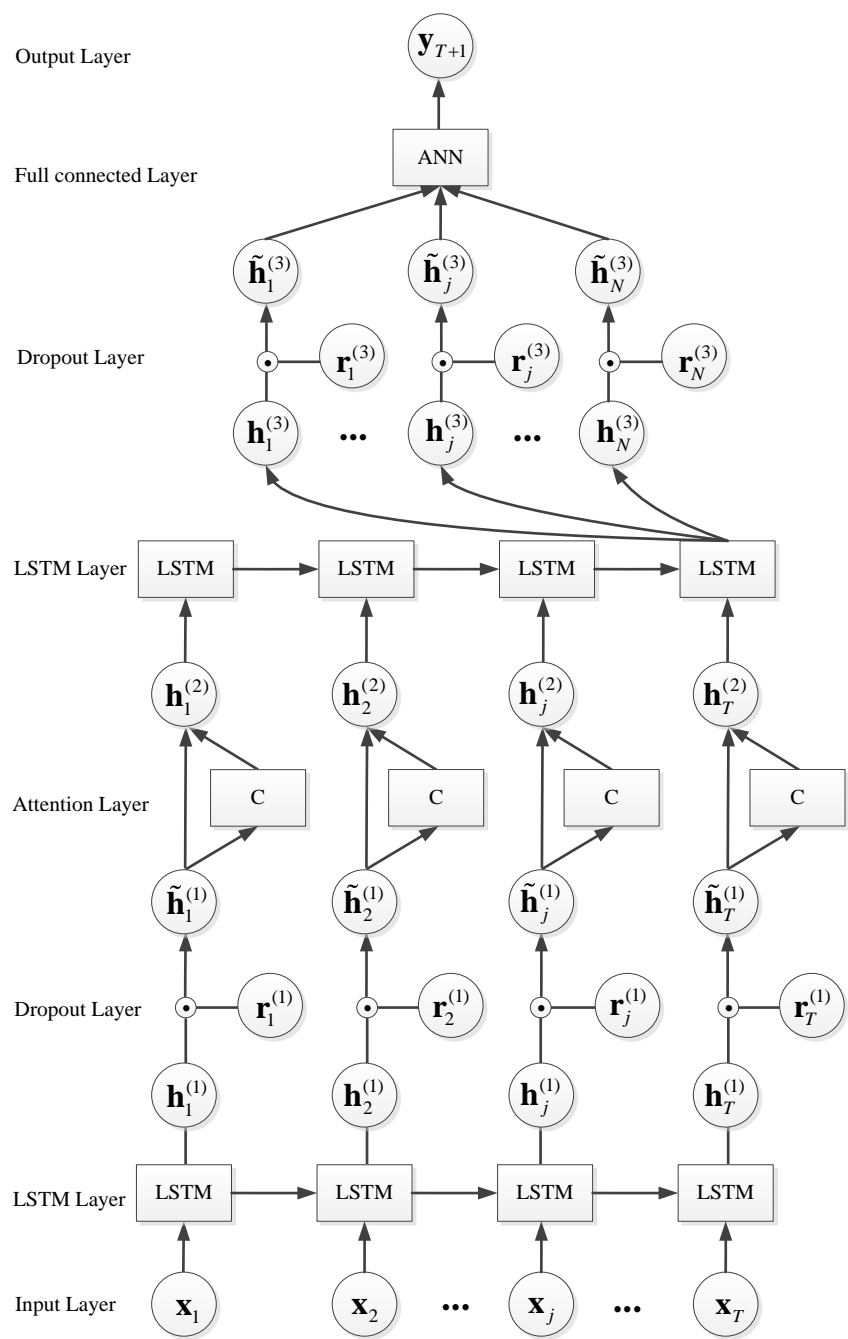

Fig.5 The framework of our attention-based LSTM forecasting system

## 3. Empirical Study
### 3.1 Data description

This article extracts data from Bloomberg on the implied volatility smile surface and options contract data of S & P500 options for each of the past ten years, from November 5, 2009 to November 5, 2019. The method of constructing Bloomberg's implied volatility smile surface is to first calculate the implied volatility discrete points of each option contract by using the BS formula based on the option contract price, and then perform arbitrage smoothing on the discrete points. Smooth volatility surface. In this process, Bloomberg uses out-of-the-money to calculate the implied volatility matrix with a moneyness greater than 100%, and uses out-of-the-money to calculate the moneyness less than The 100% implied volatility matrix is then stitched together to form a complete implied volatility matrix, which is further smoothed to obtain the implied volatility surface. In the no-arbitrage environment assumed by call-put parity, call and put options with the same expiration date and strike price often have the same implied volatility. Therefore, the implied volatility smile surface provided by Bloomberg can price both call and put options. Since the S & P500 option contract is a European option, the BS formula can be used directly in the implied volatility estimation and option pricing. Since estimating the volatility smile surface based on options contract is not an innovative work, this paper conducts deep learning prediction modeling research based on Bloomberg's estimated volatility surface, mainly to highlight the original work of this article in a limited space.

This article extracts data from Bloomberg on the implied volatility smile surface and options contract data of S & P500 options for each of the past ten years, from November 5, 2009 to November 5, 2019. The method of constructing Bloomberg's implied volatility smile surface is to first calculate the implied volatility discrete points of each option contract by using the BS formula based on the option contract price. Smooth volatility surface. In this process, Bloomberg uses out-of-the-money to calculate the implied volatility matrix with a moneyness greater than 100%, and uses out-of-the-money to calculate the moneyness less than 100% implied volatility matrix, and then the two matrices are stitched into a complete implied volatility matrix, which is further smoothed to obtain the implied volatility surface[②]. In the no-arbitrage environment with call-put parity assumptions, call and put options with the same expiration date and strike price often have the same implied volatility[③]. Therefore, the implied volatility smile surface provided by Bloomberg can price both call and put options. Since the S & P500 option contract is a European option, the BS formula can be used directly in the implied volatility estimation and option pricing. Since estimating the volatility smile surface based on options contract is not an innovative work, this paper conducts deep learning prediction modeling research based on Bloomberg's estimated volatility surface, mainly to highlight the original work of this article in a limited space.

The option implied volatility smile surface extracted from Bloomberg in this paper is shown in Figure 6. The

---

[②] The definition of the value state of the volatility smile surface is moneyness = K / F, where K is the exercise price of the option contract, and F is the price of the underlying asset corresponding to the option contract (that is, the S & P500 index price). Moneyness 100% call or put options are at-the-money. Call options with Moneyness less than 100% are in-the-money, and put options with more than 100% are in-the-money. Out-of-the-money options will become a piece of scrap paper after the expiration date, and investors will waive their right to exercise.

[③] Put-call parity refers to the basic relationship that must exist between the option price and the option price with the same expiration date. For the same underlying asset, at the same expiration date T, the call option C with the same exercise price and the put option P have a mathematical relationship: $C - P = S - Ee^{-r(T-t)}$, where r is the interest rate and t is the current time. If the equation does not hold, it means there is an opportunity for risk-free arbitrage.

x-axis represents the remaining expiration day of the option contract, the y-axis represents the value status of the option contract (moneyness), and each intersection of the x-axis and the y-axis represents a specific exercise price and expiration date. For option contracts, the z-axis represents the magnitude of the implied volatility of the contract. The volatility smile surface contains 45 discrete points, and its terms (option contract expiration dates) include March, June, December, 18, and 24 months, and the value status is 80%, 90%, 95%, 97.5%, 100%, 102.5%, 105%, 110% and 120%. Figure 6 shows that if the moneyness is close to 100%, Bloomberg provides more dense implicit volatility data because the corresponding option contract trading volume is more active. It can be seen from Figure 6 that the implied volatility of options with different expiration dates has a "smile" or "skew" phenomenon. There are dozens and hundreds of S & P500 option contracts that are actively traded every day. If you further perform interpolation fitting on the implied volatility surface, you can perform horizontal comparisons of option contracts of different types, different exercise prices, and different expiration dates. A higher point on the surface is an option contract that is more expensive, and a lower point is an option contract that is relatively cheaper.

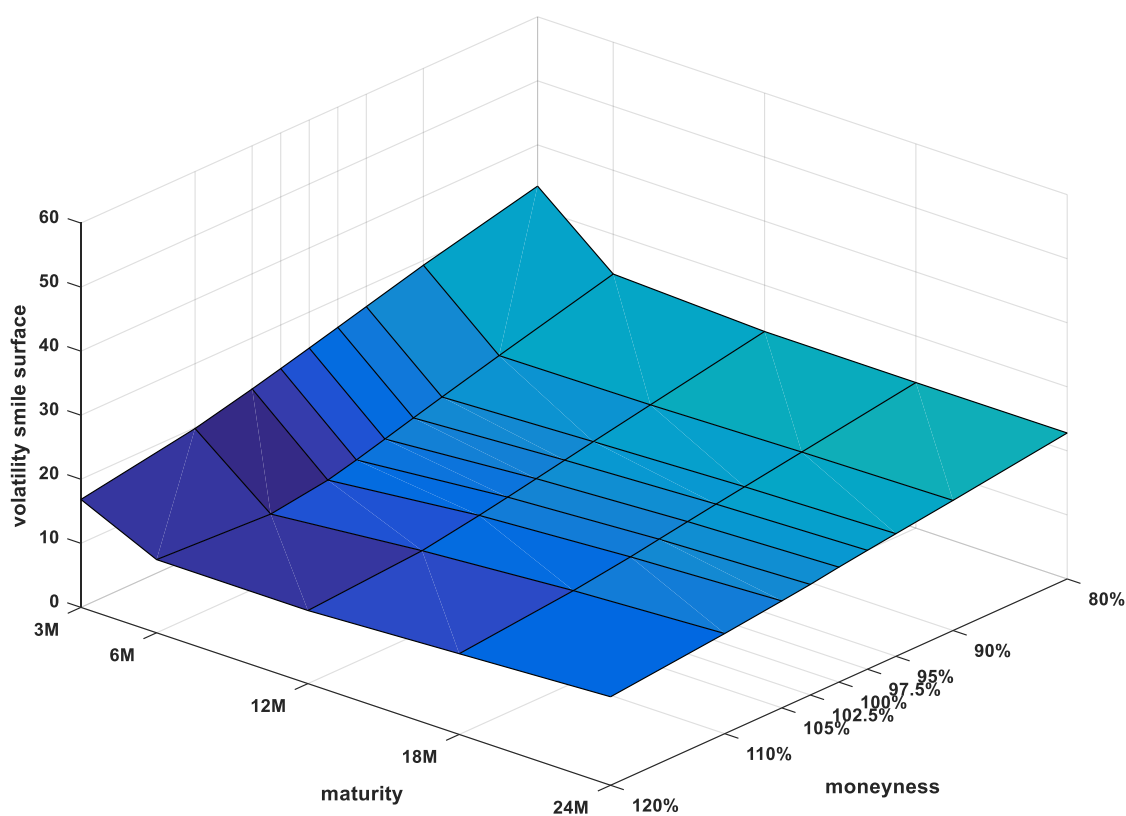

Fig.6 Implied volatility smile surface of S&P 500 index option

**3.2 Prediction of Volatility Smile Surface**

The core work of the empirical part of this paper is to use the attention-based LSTM deep learning prediction system as a tool to use the historical sequence of volatility smile surfaces as input features to predict future volatility smile surfaces. In the process of volatility surface prediction, this study is unique in that it adopts the idea of holism, that is, the volatility smile surface is directly used as an overall prediction target. This idea is similar to the study of deep learning in human motion recognition (literature). It is believed that any implicit volatility point in the volatility smile surface is not only affected by its own historical volatility sequence, but also by the impact of other historical volatility series of volatility surface. The training process of the Att-LSTM prediction system is explained in detail below, and the differences between Att-LSTM, LSTM, and MLP in static prediction and rolling prediction are compared.

### 3.2.1 Training of deep learning systems

First, this article needs to clarify the input features and output goals of the Att-LSTM deep learning system. Regarding the input characteristics of the Att-LSTM prediction system, referring to the heterogeneous investor fluctuations used in HAR-RV modeling, this article takes the implicit daily volatility of the historical daily, weekly, and monthly lines corresponding to each point of the volatility smile surface as input Features, namely:

$$x_t^d = x_t, x_t^w = \sum_t^{t-5} x_t / 5, x_t^m = \sum_t^{t-22} x_t / 5 \qquad (20)$$

The method for constructing the input features described above makes it possible to extract long-term volatility surface information from fewer input features. Compared with the previous prediction model form $\mathbf{y}_{T+1} = \text{Att-LSTM}(\mathbf{x}_T)$, the input feature $\mathbf{x}_T$ is a matrix sequence with dimensions of 45 * 3, and the prediction target $\mathbf{y}_{T+1}$ is a data matrix with dimensions of 45 * 1.

Secondly, the Att-LSTM prediction system needs to preset the parameters of the neural network structure. In this paper, the dimensions of the Att-LSTM network layer are initialized to [45,135,135,45], and the time step of the first LSTM layer is set to 3, which is used to load the daily volatility surface curve series, which is the first The shape of the LSTM layer is shape = (3,45). Regarding the Att-LSTM information transfer process presented in Figure 5, the following description will be given from the bottom to the top with specific problems: First, the input layer input layer is three 45 * 1 vectors; the first LSTM layer output vector size is 135 * 3 That is, 3 LSTM processing units are used, and the output of the previous LSTM processing unit is passed to the next LSTM processing unit in turn; the input and output of the dropout layer are 3 135 * 1 vectors; the input and output of the attention layer are also 3 135 * 1 vector, the output of the attention layer increases the activation function sofrmax; the output of the second LSTM layer is 1 135 * 1, and the network layer only allows the last LSTM processing unit to output; the LSTM processing unit will output 135 1s * 1 vector, and connect the dropout layer again to improve the learning ability. Finally, the 135 1 * 1 vectors output by the dropout layer will be further input to the fully connected network layer. After the linear activation function is calculated, the dimension will be 45 * 1 Output target.

Finally, training data and training parameters need to be specified. In this paper, the Att-LSTM model is trained in depth based on the historical 10-year data of the S & P500 option volatility smile surface. In this paper, the research sample is divided into an in-sample training set and an out-of-sample validation set, that is, the data of the first 8 years is selected as the in-sample and the remaining 2 years of the data is used as the out-of-sample. In addition, in the training parameter settings of the Att-LSTM prediction system, the optimizer is selected as rmsprop, the loss function is selected as mean_squared_error, the batch_size is set to 50, and the number of training nb_epoch is set to 400. Figure 7 shows the training error of the implied volatility smile surface prediction. The results show that the MSE of the training set within the sample and the validation set outside the sample gradually decreases and converges to a stable value with the increase of the number of trainings. The experimental results fully show that the Att-LSTM deep learning system established in this paper is very effective in predicting the volatility smile surface problem. Therefore, this conclusion lays a solid foundation for rolling prediction of the volatility smile surface using the Att-LSTM model.

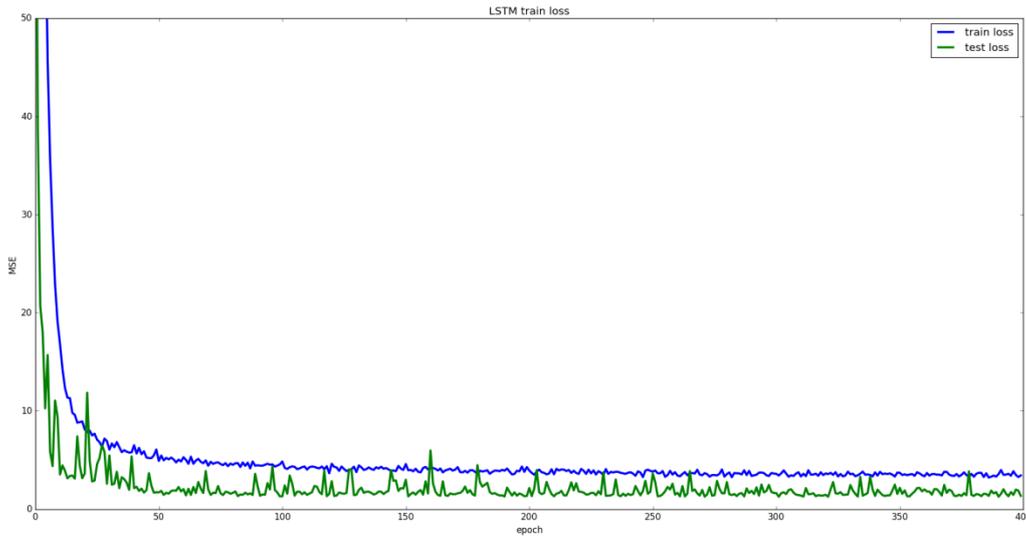

Fig.7 Training losses of implied volatility smile surface prediction

**3.2.2 Static prediction and rolling prediction**

In order to evaluate the performance of the Att-LSTM prediction system, this paper uses the same sample data and based on the same input features and prediction targets, and adds the performance comparison of LSTM prediction system and MLP prediction system. The LSTM prediction system and the MLP prediction system are designed using long and short memory networks and multi-layer perceptrons, respectively. The network structures of both include an input layer, three hidden layers, and a fully connected layer. softmax activation function. The input layers of the two prediction systems use three 45 × 1 matrix inputs, and the output of the first and second LSTM hidden layers and MLP hidden layers are three 135*1 vectors. The third hidden layer of LSTM only retains the output of the last LSTM processing unit, and the output size is a 135 * 1 vector. The output size of the third hidden layer of the MLP prediction system is an output vector of 135*1. The final fully connected layer maps the input vector of size 135*1 to the output target of 45*1 through the linear activation function linear. By comparison, the Att-LSTM prediction system adds an attention layer and a dropout layer than the LSTM prediction system, and the LSTM prediction system adds its own feedback mechanism than the MLP prediction system.

In the prediction experiment, this paper carried out static prediction and rolling prediction. The static prediction is similar to the training process of the Att-LSTM prediction system described above, and the 10-year history of the volatility smile curved surface sequence is divided into an in-sample training set for the first 8 years and an out-of-sample validation set for the next 2 years. For static prediction, this paper uses the data from the first 8 years as the training set within the sample, and performs 400 trainings on the three types of prediction systems to obtain the parameters, and then predicts the next 2 years as the validation set. Rolling prediction is based on the rolling time window for training parameters, and then using the volatility smile surface for the next 2 years as the prediction target for rolling prediction. When rolling prediction, this paper sets the rolling time window to 3 years, and uses the historical data of the volatility smile surface of the previous 3 years to train the three types of prediction systems 100 times, and rolls the trained system to predict the last 2 years. Volatility smile surface.

Since the loss functions of the three types of prediction systems are mean_squared_error, MSE itself is a key indicator for evaluating the pros and cons of the prediction system. In order to make the research results more convincing, this paper uses two other loss functions (MAE and QLIKE) to evaluate the results of static prediction and rolling prediction. The three loss functions used in this article are defined as:

$$MSE = \frac{1}{T \times M} \sum_{t=1}^{T} \sum_{m=1}^{M} (IV_{t,m} - \hat{\sigma}_{t,m})^2 \tag{21}$$

$$MAE = \frac{1}{T \times M} \sum_{t=1}^{T} \sum_{m=1}^{M} |IV_{t,m} - \hat{\sigma}_{t,m}| \tag{22}$$

$$QLIKE = \frac{1}{T \times M} \sum_{t=1}^{T} \sum_{m=1}^{M} (\log(\hat{\sigma}_{t,m}) + IV_{t,m}/\hat{\sigma}_{t,m}) \tag{23}$$

Based on the above training data and parameter settings, this paper summarizes the loss function results of the static prediction and rolling prediction of the three models into Table 1 and Table 2, respectively. By analyzing the static prediction in Table 1, it can be known that the LSTM prediction system is better than the MLP prediction system, and the Att-LSTM prediction system is better than the LSTM prediction system. Specifically, both the intra-sample prediction loss and the out-of-sample prediction loss MSE of the Att-LSTM prediction system are the smallest. Analysis of the rolling predictions in Table 2 shows that Att-LSTM's rolling prediction performance within and outside the sample is still the best, and its MSE is generally smaller than that of LSTM and MLP prediction systems. In order to analyze the intuitive performance of rolling prediction, this paper uses Figure 8 to present a set of rolling prediction performance of a series of implicit volatility with a maturity of 3 months and moneyness including 9 cases. November 5, 2019. The line "-." In Fig. 8 indicates the prediction target of the implied volatility, and the line "-" indicates the predicted value of the implied volatility. Figure 8 shows intuitively that the deviation between the predicted value of the implied volatility and the predicted target of the implied volatility is small. This performance fully illustrates that the use of deep learning to predict the implied volatility surface as a whole is theoretically scientific, Accurate in results. In summary, the Att-LSTM prediction system established in this paper has strong adaptability to the problem of volatility surface prediction.

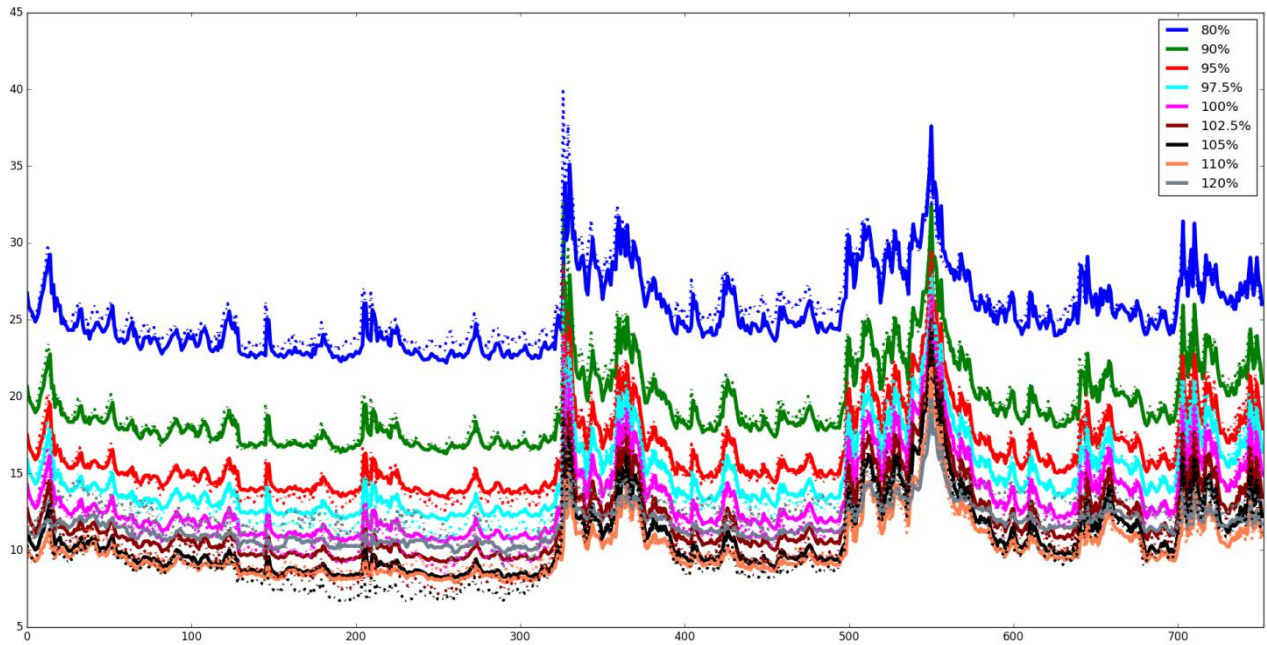

Fig.8 rolling prediction series of implied volatility

Tab.1 Non-rolling prediction performance of 3 models

| Model | In-sample | | | Out-of-sample | | |
|---|---|---|---|---|---|---|
| | MSE | MAE | QLIKE | MSE | MAE | QLIKE |
| Att-LSTM | 1.024 | 0.704 | 3.786 | 1.212 | 0.751 | 3.695 |
| LSTM | 1.325 | 1.128 | 3.828 | 1.525 | 1.165 | 4.071 |
| MLP | 1.512 | 1.259 | 4.259 | 2.367 | 1.160 | 4.510 |

Tab.2 Rolling prediction performance of 3 models

| Model | In-sample | | | Out-of-sample | | |
|---|---|---|---|---|---|---|
| | MSE | MAE | QLIKE | MSE | MAE | QLIKE |
| Att-LSTM | 0.885 | 0.666 | 3.822 | 1.083 | 0.762 | 3.641 |
| LSTM | 1.260 | 1.107 | 4.103 | 1.210 | 1.102 | 3.953 |
| MLP | 1.413 | 1.313 | 4.214 | 2.318 | 1.264 | 4.376 |

**3.3.2 Option Pricing based on Volatility Surface**

The accurate pricing of options contracts is a prerequisite for many options investment strategies. If investors want to achieve long-term and stable returns, it is difficult for investors to bear the losses caused by mispricing. According to the BSM option pricing formula, as long as the corresponding parameters are entered into the pricing model, the theoretical price of the option can be obtained. These parameters include exercise price, time to expiry date, underlying price, interest rate, and implied volatility. If it is a stock option, you also need to enter a dividend, but this article does not need to enter a dividend, because the research object is S & P500 index options. In addition to the implied volatility input, the other four input parameters are variables that can be observed from the market, so the biggest problem for investors using the BSM option pricing model is how to estimate the implied volatility more accurately.

However, in the real market, volatility is not only very difficult to determine, but a slight increase or decrease in volatility will have a huge impact on option prices. Once the implied volatility changes, the value of an option can either skyrocket or plummet. Whether it is a professional individual investor or an institutional investor who is developing a hedging strategy for a public fund, investment decisions will depend on the accuracy of his input volatility in the theoretical pricing model. The implied volatility can be derived from the market price of the options market, so the implied volatility can be the market's own prediction of future volatility. Implied volatility is the consensus reached by all traders in the market through the option bid and offer quotes for what they consider to be future volatility. It is worthwhile to say that pricing an option contract using a volatility surface is more robust than pricing an option contract using a single volatility point.

Since the S & P500 option contract is a European option contract, this article uses the BSM formula to price the option contract for two implicit volatility smile surfaces. According to the classic Black-Scholes option pricing theory, the price of an option will be affected by various factors such as the underlying asset price, the contract exercise price, the expected volatility of the underlying asset, the remaining maturity of the contract, and the risk-free interest rate. At time t, for a specific option contract, its corresponding underlying asset price, contract exercise price, and contract remaining maturity have unique values; the risk-free interest rate can refer to the US 10-year Treasury rate Given that short-term changes will not occur; the only variable that is difficult to determine in advance is volatility. If volatility can be accurately estimated, then options can be valued. The Bloomberg terminal uses the BSM formula to give an implied volatility smile surface that can be directly used for option contract pricing. The core work of this article is to use AI to further extrapolate the volatility smile surface. This article uses both surfaces to use the BSM formula to price option contracts, and lays the foundation for subsequent research on option spread strategies.

### 3.3.3 Analysis of Option Spread Strategy

The true charm of options lies in its "building blocks" nature. Investors can skillfully design many low-risk option portfolio strategies. Based on the pricing of option contracts using the implied volatility surface of options, this section designs two options combination strategies, namely intertemporal spread and butterfly options, and discusses the economic value of volatility surface prediction from the perspective of investment applications.

**(1) Time spread strategy**

This section examines the profitability of time spread strategies. Time spread [④]是 is an investment strategy that combines two options with the same underlying asset, the same strike price, and the same option type (call or put), but with different expiration dates, in order to profit from the loss of time value. The time spread strategy consists of two option contracts with the same strike price and opposite positions. When buying a forward option contract and selling a short-term option contract, it is called a time spread long; when selling a forward option contract and buying a short-term option contract, it becomes a short time spread[⑤]. Because long-term options have more time value, time spread long strategies need to pay premiums, while time spread short strategies earn premiums first.

In this section, call options are used as an example to construct a time spread long portfolio, and to explore the profit and loss of time spreads at different index prices, as shown in Table 4. 1) Assuming the exercise price of the two options is K, we sell the short-term options at the premium C1 and buy the forward options at the premium C2. At this time, the cost of constructing a long portfolio of calendar spreads is C1-C2.[⑥] 2) When the

---

[④] The time spread is also called "calendar spread".
[⑤] The position ratio of the time spread strategy is usually 1: 1, but investors can change the position ratio of the time spread based on the judgment of bear market, bull market and market neutrality.
[⑥] Because forward options have a higher time value than short-term options, forward option premiums at the same strike price are higher than short-term options, that is, C2> C1.

short-term option expires, if the underlying asset price St ⩽ K, the short-term call option with an exercise price of K will be abandoned and its contract value decayed to zero, at which time the short premium C1 can be retained; The forward call option still has time value, but its value will decay to Ct2 (Ct2 <C2). At this time, the purchased forward call option will also lose C2-Ct2; in this case, the profit and loss of the long spread combination is C1 + Ct2-C2. 3) When the short-term call option expires, if the price of the underlying asset St> K, the value of the short-term call option will increase from C1 to C1 + (St-K), and the short position of the short-term call option Will be exercised, the resulting loss is St-K; the value of the forward call option contract will also increase from C2 to Ct2 (Ct2> C2), and the long call option will increase the profit of Ct2-C2; therefore, in this case The profit and loss of the long spread combination is C1- (St-K) + Ct2-C2.

As shown in Figure 9, when the price of the underlying asset is close to K, the time spread combination can earn a profit of C1 + Ct2-C2, and a loss will occur if it deviates far from the exercise price K. The time spread strategy has two sources of profit: On the one hand, the profit and loss of the time spread strategy involves the time value decay characteristics of the option. When the short-term call option approaches the expiration date, its time value decays faster than the long-term call option. . It is because of how quickly the time value of short-term and forward options decays, creating a time spread that allows investors to profit from it (such as the case where the underlying price is the exercise price K). On the other hand, if the target asset is consolidated in the future, the price of the underlying asset can be used as the exercise price to build a neutral calendar option strategy; if the future is predicted to rise, a higher exercise price can be set to construct a bull calendar option strategy; If you predict the future decline, you can set a lower strike price to build a bear market calendar option strategy. Therefore, the time spread strategy can not only benefit from the passage of time value, but also benefit from large fluctuations in the price of the underlying asset.

Tab.4 Profit of long calendar spread using put options

| Index price change | Profit of short-term call option | Profit of long-term call option | Profit of calendar spread portfolio |
|---|---|---|---|
| $S_t \leq K$ | $C_1$ | $C_{t2}-C_2$ | $C_1+C_{t2}-C_2$ |
| $S_t > K$ | $C_1-(S_t-K)$ | $C_{t2}-C_2$ | $C_1-(S_t-K)+C_{t2}-C_2$ |

In the process of concrete strategy construction, this paper uses the volatility prediction of the underlying asset S & P500, including whether the actual volatility is used to predict whether the underlying asset is consolidating or trending. In addition, in the exit design of the time spread strategy, this paper does not mechanically wait for the short-term option contract to expire before it ends. Instead, it uses a mobile stop loss and profit to dynamically track the risk or profit of the time spread strategy. Once the take profit or stop loss signal is started, the transaction is ended early. After pricing the BSM option contract using the Bloomberg volatility surface and the AI-predicted volatility surface, combined with the above-mentioned time spread strategy design scheme, this paper obtains the strategic return curve of Figure 10. From Figure 10, it can be seen that the time spread strategy using the implicit volatility smile curved surface predicted by AI has a higher return and Sharpe ratio (For more details, please contact the author to discuss).

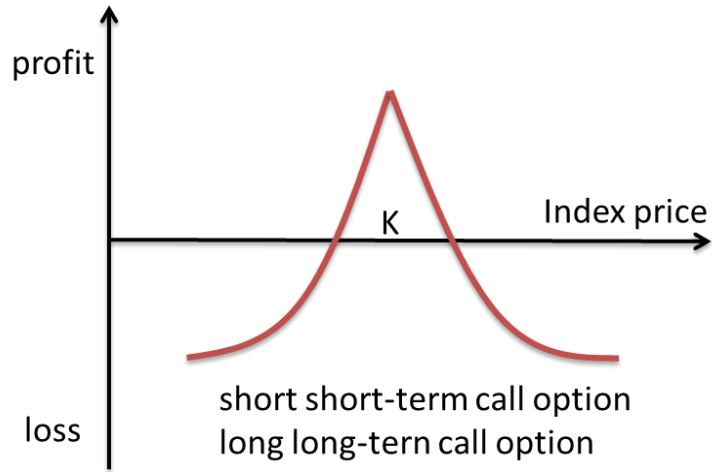

Fig.9 Profit of long calendar spread using call options

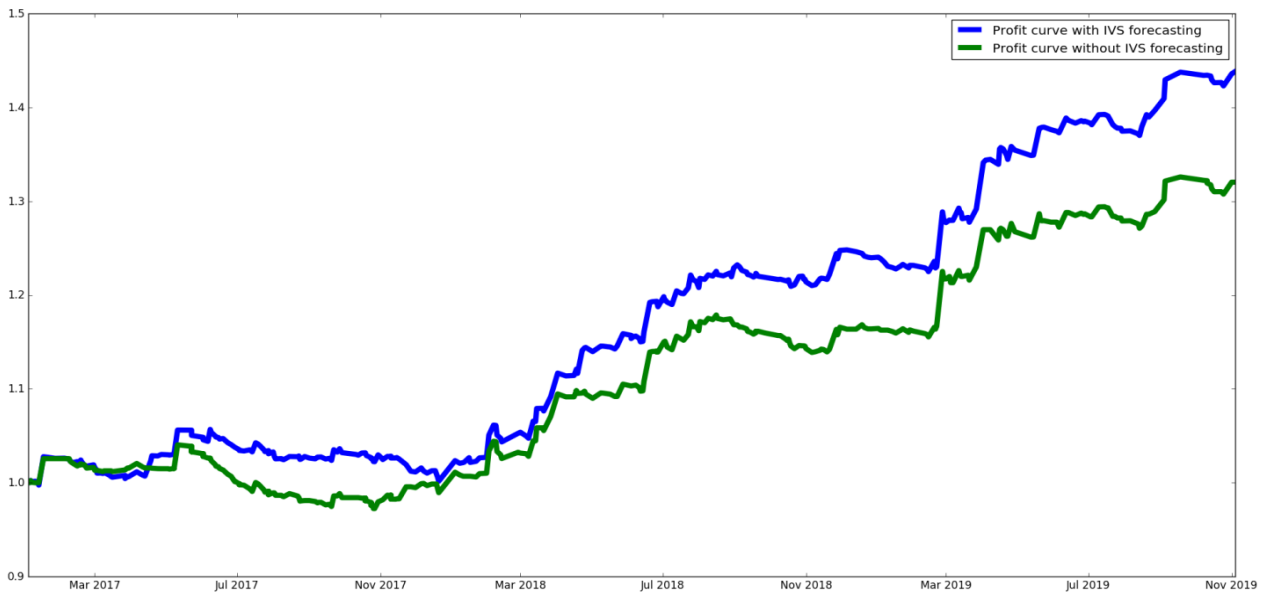

Fig.10 Profit curve of calendar spread based on BSM pricing

**(2) Butterfly spread strategy**

This section examines the profitability of the butterfly spread strategy. The time spread strategy involves two option contracts, while the butterfly spread strategy will involve three option contracts. The butterfly spread strategy is to make use of unreasonable spreads between three option contracts with the same expiration date and different strike prices for combined transactions. It consists of two sets of option spreads in opposite directions that share the middle strike price. The biggest feature of the butterfly spread combination is that it includes a combination of three options of the same contract type (both call or put options), the same expiration time, and the same exercise spread. The relationship between the exercise price of the three option contracts satisfies: K1 <K2 <K3 and K1 + K3 = 2K2. In the long butterfly spread combination, each buys an option with an exercise price of K1 and K3, and sells two options with an exercise price of K2. The short butterfly spread combination operates in the opposite direction, selling one option at the strike price of K1 and K3, and buying two options at the strike price of K2. Therefore, the position composition ratio of the butterfly spread combination can be abbreviated as 1: 2: 1.

This section uses a bullish butterfly spread strategy as an example to discuss the losses or gains of the butterfly spread combination under different price changes of the underlying asset, as shown in Table 3. 1) Assuming the three option contract options are C1, C2, and C3, constructing a long butterfly spread strategy requires payment of a premium of 2C2-C1-C3, and building a short butterfly spread strategy (short butterfly) earns first Take the same royalties. 2) When the price of the underlying asset is less than K1, investors who buy three call options will give up exercise, so the profit and loss of the long butterfly strategy of the call option is 2C2-C1-C3. 3) When the price of the underlying asset is in the range [K1, K2], call options with strike prices K2 and K3 give up exercise, and the combined profit and loss of the long-butterfly strategy is St-K1 + 2C2-C1-C3. 4) When the underlying asset price is in the interval [K2, K3 ], Call options with an exercise price of K3 give up exercise, and the combined profit and loss of the long-butterfly strategy is K3-St + 2C2-C1-C3. 5) When the target price is greater than K3, the combined profit and loss of the long-butterfly strategy is 2C2 -C1-C3.

As shown in Figure 11, for a multi-head butterfly strategy, when the target price is the exercise price K2, the combination can achieve the maximum return; when the price of the underlying asset deviates from the exercise price K2, the combination becomes increasingly valueless loss is the premium paid. For the short butterfly strategy, the opposite is true. The farther the target price deviates from the exercise price K2, the better. The maximum return is the initial premium earned, and the maximum loss occurs when the asset price is K2. Therefore, when the index price is close to the execution price K2, the long butterfly strategy has a positive return; when the index price deviates from the execution price K2, the short butterfly strategy also has a large return. In practical applications, when the underlying index is in a narrow fluctuation range, investors can buy butterfly options to obtain stable returns. If an investor determines that an event has a significant impact on the market and the index volatility increases, but the direction is not clear, then investors can sell butterfly options to obtain positive returns. Therefore, the biggest feature of the butterfly spread combination is that it has only limited risks and is also a typical volatility speculation strategy.

Tab.3 Profit of a put option butterfly strategy

| Index price change | Profit of long call option of K1 | Profit of short call option of K2 | Profit of long call option of K3 | Profit of calendar butterfly portfolio |
|---|---|---|---|---|
| $S_t \leq K_1$ | $-C_1$ | $2C_2$ | $-C_3$ | $2C_2-C_1-C_3$ |
| $K_1 < S_t \leq K_2$ | $S_t-K_1-C_1$ | $2C_2$ | $-C_3$ | $S_t-K_1+2C_2-C_1-C_3$ |
| $K_2 < S_t \leq K_3$ | $S_t-K_1-C_1$ | $2C_2-2(S_t-K_2)$ | $-C_3$ | $K_3-S_t+2C_2-C_1-C_3$ |
| $S_t > K_3$ | $S_t-K_1-C_1$ | $2C_2-2(S_t-K_2)$ | $S_t-K_3-C_3$ | $2C_2-C_1-C_3$ |

In the process of concrete strategy construction, similar to the time spread strategy, this section also uses the volatility prediction of the underlying asset S & P500, which includes the use of true volatility to predict whether the underlying asset is consolidation or trend. In the exit design of the time spread strategy, this section does not mechanically wait for the option contract to expire before it ends. It also uses a mobile stop loss and profit to dynamically track the risk or profit of the time spread strategy. Once the take profit or stop loss signal is started, the transaction is ended early. After using Bloomberg's volatility surface and AI-predicted volatility surface to price BSM option contracts, combined with the above-mentioned time spread strategy design scheme, this paper obtains the strategic return curve of Figure 12. It can be seen from Figure 12 that the time spread strategy using the implicit volatility smile curved surface predicted by AI has a higher yield and Sharpe ratio (Contact the author for more details).

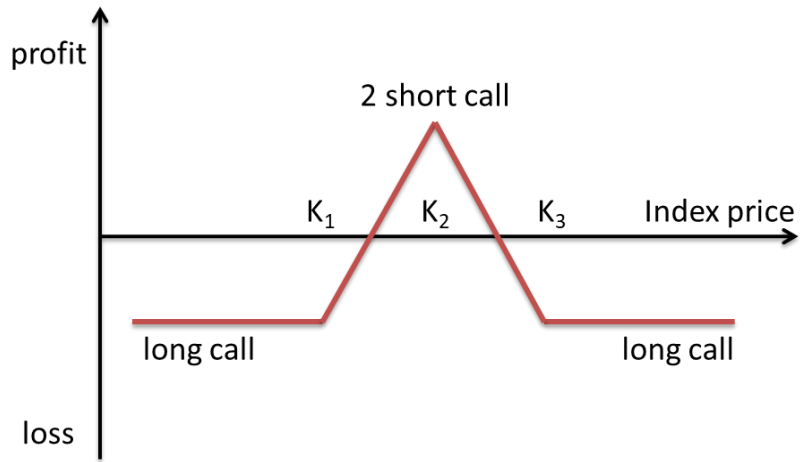

**Fig.11 Profit of long butterfly using call options**

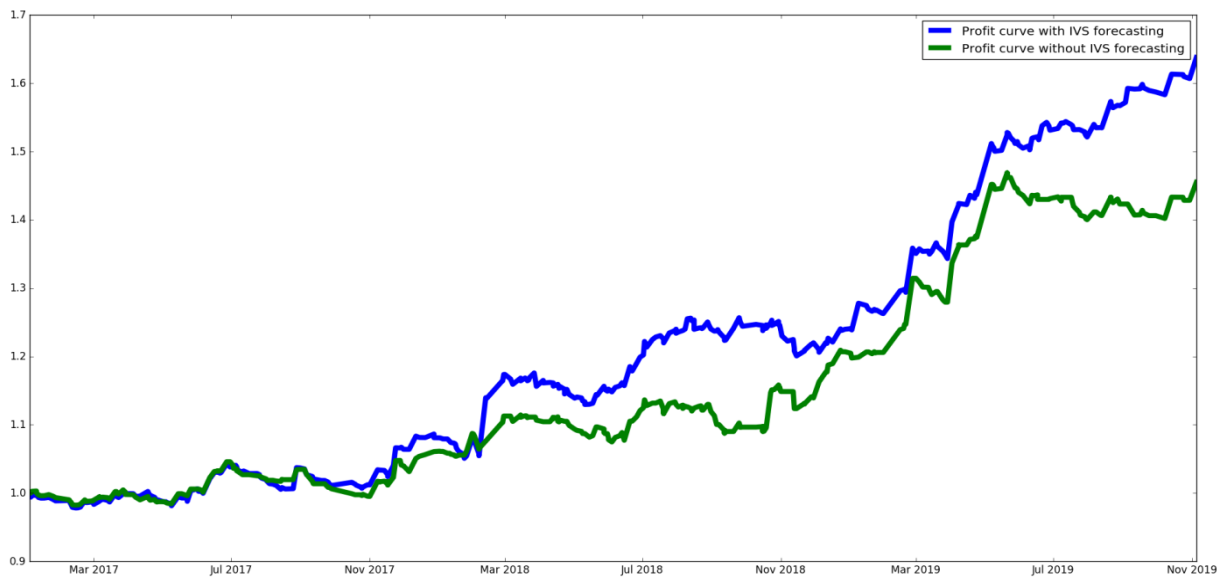

**Fig.12 Profit curve from butterfly Arbitrage based on BSM pricing**

## 4. Conclusions

    The implied volatility smile surface is the basis of option pricing, and the dynamic evolution of the option volatility smile surface is difficult to predict. In this paper, attention mechanism is introduced into LSTM, and a volatility surface prediction method combining deep learning and attention mechanism is pioneeringly established. LSTM's forgetting gate makes it have strong generalization ability, and its feedback structure enables it to characterize the long memory of financial volatility. The application of attention mechanism in LSTM networks can significantly enhance the ability of LSTM networks to select input features. This paper considers the discrete points of the implied volatility smile surface as an overall prediction target, extracts the daily, weekly, and monthly option implied volatility as input features and establishes a set of LSTM-Attention deep learning systems. Using the dropout mechanism in training reduces the risk of overfitting. For the prediction results, we use arbitrage-free smoothing to form the final implied volatility smile surface. This article uses the S&P 500 option market to conduct an empirical study. The research shows that the error curve of the LSTM-attention prediction system converges, and the prediction of the implied volatility surface is more accurate than other predicting system. According to the implied volatility surface of the 3-year rolling forecast, the BS formula is used to pricing the option contract, and then a time spread strategy and a butterfly spread strategy are constructed respectively. The experimental results

show that the two strategies constructed using the predicted implied volatility surfaces have higher returns and Sharpe ratios than that the volatility surfaces are not predicted. This paper confirms that the use of AI to predict the implied volatility surface has theoretical and economic value. The research method provides a valuable reference for option pricing and strategy using deep learning.